\documentclass[]{elsart}
\usepackage[]{graphicx}
\graphicspath{{figures/}}
\usepackage{amsmath}
\DeclareMathOperator{\Li}{Li}

\begin{document}
\begin{frontmatter}
\begin{flushleft}
TTP09-17\\
SFB/CPP-09-50\\
arXiv:0907.2120
\end{flushleft}
\title{Reconstruction of heavy quark current correlators at ${\cal O}(\alpha_s^3)$} 
\author[Karlsruhe]{Y. Kiyo}, 
\author[Karlsruhe]{A. Maier}, 
\author[Karlsruhe]{P. Maierh\"ofer},
\author[Karlsruhe]{P. Marquard}
\address[Karlsruhe]{Institut f\"ur Theoretische Teilchenphysik, 
Universit\"at Karlsruhe, Karlsruhe~Institute~of~Technology~(KIT), 76128 Karlsruhe, Germany}
\begin{abstract}
We construct approximate formulas for the ${\cal O}(\alpha_s^3)$ QCD contributions to
vector, axial-vector, scalar and pseudo-scalar quark current correlators, which are valid for
arbitrary values of momenta and masses. The derivation 
is based on conformal mapping and the Pad\'e approximation procedure and
incorporates known expansions in the low energy, threshold and high energy
regions. We use our results to estimate additional terms in 
these expansions.
\end{abstract}

\end{frontmatter}

\section{Introduction}
\label{sec:intro}
Correlators of heavy quark currents in different kinematical regions are of interest for a number of
 phenomenological applications. These two-point functions only depend on
 two scales, namely the square of the external four-momentum $q^2$ and
 the heavy quark mass $m$. Many of the applications focus on one of
 three distinct kinematical regions: The low energy region with $q^2
 \approx 0$, the quark pair production threshold at $q^2=4m^2$ and the
 (euclidean) high energy region $-q^2 \to \infty$. 

Moments in the low energy expansion can be used for
 precise extractions of charm and bottom quark masses via sum rules
 \cite{Novikov:1977dq} (for reviews see \cite{Reinders:1984sr,Colangelo:2000dp,Ioffe:2005ym}),
 whereas threshold and high energy expansions are directly related to
 production cross sections for $t \bar{t}$, 
charmed hadrons or bottom hadrons in the respective energy regions.

The aforementioned expansions for the correlators in the three regions are known relatively well: In the low energy
region at ${\cal O} (\alpha_s^2)$ the leading eight coefficients were
 computed more than ten years ago \cite{Chetyrkin:1995ii,Chetyrkin:1996cf,Chetyrkin:1997mb}, and as of
 today as many as $30$ moments are known \cite{Boughezal:2006uu,Maier:2007yn}.
At ${\cal O}(\alpha_s^3)$, however,  only the first physical moment
proportional to $(q^2)^1$ for
axial-vector and scalar correlators \cite{Sturm:2008eb} and the first and
second moment for the pseudo-scalar and vector correlators \cite{Sturm:2008eb,Boughezal:2006px,Chetyrkin:2006xg,Maier:2008he} have been available until
recently. The third moment for all four current correlators and the
fourth moment for the pseudo-scalar correlator are computed in Ref. \cite{Maier:2009}.

Threshold expansions for correlators are expansions in the small heavy
quark velocity $v=\sqrt{1-4m^2/q^2} \ll 1$. Currently all of the
necessary machinery for NNLO threshold expansion is known (see for
instance Ref. \cite{Hoang:2000yr} and references therein). This
means, that all terms of order
$(\alpha_s^n/v^{n-1})\cdot\{1,\,v,\,v^2\}$ are in principle
known for all correlators. Explicit expansions for the axial-vector,
scalar and vector correlators can be derived from Refs.
\cite{Penin:1998ik,Beneke:1999fe,Pineda:2006ri} (see Appendix \ref{sec:thr_exp}).

For high energies the leading five coefficients in the expansion are
known  at ${\cal O}(\alpha_s^2)$ for the case of scalar and pseudo-scalar
currents \cite{Harlander:1997xa} while for the
vector and axial-vector currents seven terms are available \cite{Chetyrkin:1997qi,Harlander:1997kw}.
At order $\alpha_s^3$ the first two terms in the high energy expansion of
the vector correlator have been published in Refs. \cite{Baikov:2004ku,Baikov:2009uw}.
More information is available for the absorptive parts of the
correlators, which  correspond to the logarithmic terms
in the high energy expansions. Here the first three
coefficients are known for the vector and axial-vector correlators
\cite{Gorishnii:1991vf,Surguladze:1991tg,Chetyrkin:1990kr,Chetyrkin:1996hm,Chetyrkin:2000zk}.
The leading term in the absorptive part has been computed recently 
for the vector and scalar correlators \cite{Baikov:2009uw,Baikov:2008jh,Baikov:2005rw} even at order $\alpha_s^4$.

Still, it would be desirable to have results
for the correlators which are valid for arbitrary energies in addition
to these expansions. One obvious
benefit would be the possibility to expand such a result in a
kinematical region of interest in order to obtain even more coefficients in
the expansion. It would also be possible to predict values of cross sections for
intermediate regions between threshold and high energies, where the mere
expansions may not be very accurate anymore. Last but not least, the
full energy dependence is essential for those QCD sum rule approaches to
quark mass determination which use either the Borel transformation of
the correlator \cite{Novikov:1977dq} or so-called contour-improved
perturbation theory \cite{Pivovarov:1991bi,Pivovarov:1991rh,LeDiberder:1992te}.

Unfortunately, analytical results which are valid for arbitrary energies
only exist up to ${\cal O}(\alpha_s)$ \cite{Kallen:1955fb}. Still, it is possible to
reconstruct the full energy dependence approximately from the known
expansions at higher orders. Using a seminumerical method based on
Pad\'e approximations
\cite{Baker:1975,Broadhurst:1993mw,Fleischer:1994ef,Broadhurst:1994qj,Baikov:1995ui},
the correlators of all four currents were reconstructed at  ${\cal
  O}(\alpha_s^2)$ \cite{Chetyrkin:1995ii,Chetyrkin:1997mb,Chetyrkin:1998ix}. Moreover, it
was demonstrated in \cite{Hoang:2008qy} that in spite of the
rather low amount of information available at  ${\cal O}(\alpha_s^3)$ it
is still viable to reconstruct the vector correlator and predict
expansion coefficients with decent accuracy. Recent computations of
additional terms in the expansions in the low and
high energy domains \cite{Maier:2008he,Maier:2009} have confirmed these predictions and render the
application of the approximation procedure to other correlators feasible.

In this work we use the Pad\'e approximation method to reconstruct the
vector, axial-vector, scalar and pseudo-scalar heavy quark current
correlators at ${\cal O}(\alpha_s^3)$ and derive approximations to previously unknown expansion coefficients in the low energy,
threshold and high energy regions.

This paper is structured in the following way:
In Section \ref{sec:meth} we set up our conventions and explain the
general approximation method. Section \ref{sec:calc} contains details of
the application of the method to the different correlators and the
choice of physically meaningful approximants. The results for the
reconstructed correlators and the estimates for new expansion
coefficients can be found in Section \ref{sec:res}. In Section
\ref{sec:conc} we summarise the work and give our conclusions.

\section{Methods}
\label{sec:meth}

\subsection{Polarisation functions}
\label{sec:polf}

It is convenient to explicitly extract the Lorentz structure of the
heavy quark current correlators and define polarisation functions 
$\Pi^\delta(q^2),\,\Pi_L^\delta(q^2)$:
\begin{align}
  \label{eq:1}
   (-q^2 g_{\mu\nu} + q_\mu q_\nu) \Pi^\delta (q^2) + q_\mu q_\nu
   \Pi_L^\delta(q^2) =& i\int dx e^{iqx} \langle 0 |Tj_\mu^\delta(x)
   j_\nu^\delta(0)|0\rangle \\
   &\hspace{12ex}\mbox{for} \quad \delta=v,a, \notag \\
   q^2 \Pi^\delta(q^2) =& i \int dx e^{iqx} \langle 0 |Tj^\delta(x)
   j^\delta(0)|0\rangle \\
   &\hspace{12ex}\mbox{for} \quad \delta=s,p, \notag
\end{align}
where the currents are defined as
\begin{align}
  j_\mu^v = \bar\psi \gamma_\mu \psi, \quad j_\mu^a = \bar\psi
  \gamma_\mu \gamma_5 \psi, \quad j^s = \bar \psi\psi, \quad j^p =
  i \bar\psi\gamma_5 \psi. 
\end{align}
The longitudinal polarisation functions $\Pi_L^\delta(q^2)$ will not be considered any
further in this work: $\Pi_L^a(q^2)$ is closely related to the pseudo-scalar
polarisation function $\Pi^p(q^2)$ by a Ward identity and $\Pi_L^v(q^2)$ even vanishes identically.
We do not take into account singlet contributions originating from
diagrams with massless cuts\footnote{The ${\cal
    O}(\alpha_s^2)$ singlet contributions and Pad\'e approximations to
  them are discussed in
  Ref. \cite{Chetyrkin:1998ix}.}
 and choose the normalisation
\begin{equation}
  \label{eq:2}
  \Pi^\delta(0)=0\,.
\end{equation}
The perturbative expansions of the polarisation functions read
\begin{equation}
  \label{eq:3}
  \Pi^\delta = \Pi^{(0),\delta} + C_F
  \Pi^{(1),\delta} \frac{\alpha_s}{\pi}+ 
  \Pi^{(2),\delta} \left(\frac{\alpha_s}{\pi}\right)^2+  \Pi^{(3),\delta} \left(\frac{\alpha_s}{\pi}\right)^3+ \dots\,,
\end{equation}
where $C_F=\frac{4}{3}$ is the quadratic Casimir operator for the adjoint representation.
A natural variable to describe the behaviour of $\Pi^\delta$ is given by
\begin{equation}
  \label{eq:4}
  z = \frac{q^2}{4m^2}\,,
\end{equation}
where $m$ denotes the heavy quark mass defined in the on-shell scheme.

For the construction of approximants we need expansions in the low
energy, threshold and euclidean high energy regions, which correspond to
$z=0,\,1\text{ and } -\infty$ respectively. Around $z=0$ the expansion reads
\begin{equation}
  \label{eq:5}
  \Pi^{(i),\delta}(z) = \frac{3}{16\pi^2}\sum_{n=1}^\infty C^{(i),\delta}_n z^n\,.
\end{equation}
The renormalisation scale $\mu$ is set to $m$, so that the coefficients
$C^{(i),\delta}_n$ are simply real numbers. 

Around $z=1$ we have
\begin{equation}
  \label{eq:6}
  \Pi^{(i),\delta}(z) = \sum_{k=k_0}^\infty \sum_{l\geq 0}
  K^{(i),\delta}_{\frac{k}{2}\,l}
  (1-z)^\frac{k}{2} \log^l (1-z)\,.
\end{equation}
The lower bound $k_0$ of the sum is $3-i$ for the axial-vector and
scalar correlators and $1-i$ for the vector and pseudo-scalar
correlators. In the na\"ive Taylor series we omit
the second index, i.e. $ K^{(i),\delta}_\frac{k}{2} \equiv K^{(i),\delta}_{\frac{k}{2}\,0}$

Finally, for $z \to -\infty$ we define
\begin{equation}
  \label{eq:7}
  \Pi^{(i),\delta}(z) = \sum_{n=0}^\infty \sum_{l\geq 0}
  D^{(i),\delta}_{n\,l} \left(\frac{1}{z}\right)^n \log^l(-4z)\,,
\end{equation}
where we again use the convention $ D^{(i),\delta}_n \equiv  D^{(i),\delta}_{n\,0}$.

\subsection{Pad\'e approximation}
\label{sec:pade1}
The Pad\'e approximant $p_{n,m}(x)$ to a function $f$ is defined as
\begin{equation}
  \label{eq:8}
  p_{n,m}(x) =  \frac{\sum_{i=0}^n a_i x^i}{1+\sum_{i=1}^m b_i x^i}\,.
\end{equation}
The coefficients $a_i,\,b_i$ are usually fixed by imposing $n+m+1$ conditions of the form
\begin{equation}
  \label{eq:9}
  p_{n,m}^{(j)}(x_i)=f^{(j)}(x_i)\,,
\end{equation}
where $f^{(j)}$ is the $j$-th derivative of $f$. As long as all
constraints have this form, a unique solution for the coefficients
$a_i,\,b_i$ is guaranteed \cite{Baker:1975}.

A na\"ive application of the Pad\'e approximation method will, however, fail
for the functions $\Pi^{(i),\delta}(z)$ 
because contrary to the Pad\'e approximants (Eq. (\ref{eq:8})) they are
not meromorphic everywhere in the complex plane. There are two major
aspects of this problem, which have to be considered:
\begin{itemize}
\item 
   The functions $\Pi^{(i),\delta}(z)$ diverge logarithmically for $z \to -\infty$. There
  are also logarithmic contributions at threshold.
\item There is a branch cut along the real axis starting from $z=1$.
\end{itemize}
This behaviour can obviously not be reproduced accurately by a Pad\'e approximation.

The first problem related to the appearance of logarithms can be cured by splitting $\Pi^{(i),\delta}(z)$
into two parts,
\begin{equation}
  \label{eq:10}
  \Pi^{(i),\delta}(z) = \Pi^{(i),\delta}_{reg}(z) + \Pi^{(i),\delta}_{log}(z)\,,
\end{equation}
so that $\Pi^{(i),\delta}_{log}(z)$ is a suitable function containing
all known logarithmic contributions to $\Pi^{(i),\delta}(z)$.  
In this way the problem reduces to finding an approximation to
$\Pi^{(i),\delta}_{reg}(z)$. 

In the case of the vector and axial-vector polarisation functions, 
the functions $\Pi^{(i),\delta}(z)$ with $i>1$ also show the well-known Coulomb
singularity at threshold in addition to the logarithmic behaviour. Being
proportional to $\left(1-z\right)^{(1-i)/2}$ this singularity is, however,
 meromorphic and can be  incorporated into the construction of
 the Pad\'e approximants.

\begin{figure}
$\begin{array}{ccc}
\includegraphics*[width =.99\linewidth]{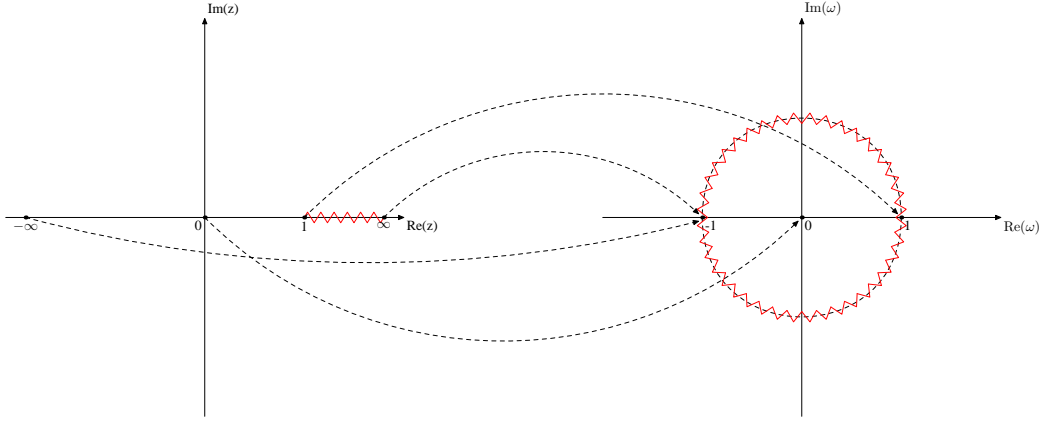}
\end{array}$
\caption{Conformal mapping of the complex plane onto the unit
  circle. The points $z=0$ and $z=1$ correspond to $\omega=0$ and
  $\omega=1$, respectively. $z\pm \infty$ goes to $\omega=-1$. The
  branch cut starting at $z=1$ is mapped onto the perimeter of the circle. } 
\label{fig:1}
\end{figure}

The second problem related to the branch cut is treated in a different
way: We map the complex plane (including its cut) onto the unit circle in such a way that the
branch cut is mapped onto the perimeter (see Fig.~\ref{fig:1}). This
can be achieved via the conformal transformation  
\begin{equation}
  \label{eq:11}
  z \to \frac{4\omega}{(1+\omega)^2}\,.
\end{equation}
The functions $\Pi^{(i),\delta}_{reg}(\omega)$ are now suitable for the
Pad\'e approximation procedure.

In the next step we  fix the coefficients $a_i,\,b_i$ in
Eq. \eqref{eq:8}. From the expansions in the low energy, threshold and
euclidean high energy regions we know the values of the $\Pi^{(i),\delta}(z)$ and
some of their derivatives for $z=0$, $z=1$, and $z \to
- \infty$. While the information from the first two regions can be used
directly to fix the derivatives of $\Pi^{(i),\delta}_{reg}(\omega)$ at
$\omega=0$ and $\omega=1$, respectively, the expansion around $z \to-
\infty$ is quite different from the expansion around $\omega=-1$.

In order to also incorporate the information
from the high energy region one can use the following auxiliary function
\cite{Chetyrkin:1998ix}
\begin{equation}
  \label{eq:12}
  P_n(\omega) = 
  \frac{(4\omega)^{n-1}}{(1+\omega)^{2n}}\left(\Pi^{(i),\delta}_{reg}(\omega)-\sum_{j=0}^{n-1}\frac{(1+\omega)^{2j}}{(4\omega)^j} \frac{1}{j!} \left(\frac{d}{d(1/z)}\right)^j \Pi^{(i),\delta}_{reg}(z)\Big|_{z=-\infty}\right)\,,
\end{equation}
where $n$ is the highest known power of $1/z$ in the high
energy expansion. $P_n(-1)$ corresponds to the coefficient of $1/z^n$,
while all other terms in the expansion of $\Pi^{(i),\delta}_{reg}(z)$
around $z\to - \infty$ together with the terms from the low energy expansion
determine the behaviour of $P_n(\omega)$ around $\omega=0$. If the
expansion around $z=0$ is available up to $z^m$ the values 
$P_n(0), P_n'(0), \dots,P_n^{(n+m-1)}(0)$ together with $P_n(-1)$
can be used for the construction of Pad\'e approximants. Additional
knowledge about the $l$ leading terms in the threshold expansion
translates into $P_n(1), P_n'(1), P_n''(1),\dots,P_n^{(l-1)}(1)$.

Note that the construction Eq. \eqref{eq:12} in general produces
half-integer powers of $1/z$ in the high energy expansion of
the reconstructed function $\Pi^{(i),\delta}_{reg}(z)$. Since only terms
with integer powers of $1/z$ may appear in the expansion, we explicitely
require that the terms with half-integer powers vanish.
These additional constraints do not have the form of Eq. \eqref{eq:9}. 
As a result, the solution to the system of equations
which determines the coefficients of the Pad\'e approximant is, in general, no longer
unique.

The Coulomb singularity proportional to $\left(1-z\right)^{(1-i)/2}$, which
appears in the vector and the axial-vector correlator, can easily be
incorporated by defining $P_n(\omega)$ with an additional factor
$(1-\omega)^{1-i}$ on the right hand side of Eq. \eqref{eq:12} so
that the limit $\omega \to 1$ becomes regular. 

\section{Reconstruction of polarisation functions}
\label{sec:calc}

In this Section we explain the details of our calculation and list the
input from known expansions.

\subsection{Subtractions}
\label{sec:subtractions}

As explained in Section \ref{sec:pade1}, the first step consists of
absorbing the logarithmic contributions in the high energy and threshold
expansions into a function $\Pi^{(3),\delta}_{log}(z)$. This function must
be chosen carefully in order not to introduce undesired additional
singularities in $\Pi^{(3),\delta}_{reg}(z)$. It is very convenient to use lower order analytical results
as auxiliary functions \cite{Hoang:2008qy,Baikov:1995ui}, namely
\begin{align}
  \label{eq:13}
  \Pi^{(0),v}(z)=&\frac{3}{16 \pi^2}\left(\frac{20}{9}+\frac{4}{3z}-\frac{4(1-z)(1+2z)}{3z}G(z)\right)\,,\\
\Pi^{(1),v}(z)=&\frac{3}{16 \pi^2}\left[\frac{5}{6}
+\frac{13}{6z}-(1-z)\frac{3+2z}{z}G(z)+(1-z)\frac{1-16z}{6z}G(z)^2 \right.\notag\\
&\left. -\frac{1+2z}{6z} \left(1+2z(1-z)\frac{d}{dz}\right)\frac{I(z)}{z}\right]\,,
\end{align}
with
\begin{align}
  \label{eq:14}
 I(z)=& 6\big(\zeta_3+4\Li_3(-u)+2\Li_3(u)\big)
 -8\big(2\Li_2(-u)+\Li_2(u)\big)\log\left(u\right)
 \notag\\&-2\big(2\log\left(1+u\right)+\log\left(1-u\right)\big)\log\left(u\right)^2\,,\\
G(z) =& \frac{1}{2z}\frac{\log\left(u\right)}{\sqrt{1-\frac{1}{z}}}\,,
\hspace{3cm} u = \frac{\sqrt{1-\frac{1}{z}}-1}{\sqrt{1-\frac{1}{z}}+1}\,. 
\end{align}
First, we treat the logarithmic behaviour at threshold. The relevant
expansions of the four correlators read
\begin{align}
  \label{eq:15}
\Pi^{(3),v}(z)=&\frac{2.63641}{1-z}+\frac{-27.2677+0.678207n_l}{\sqrt{1-z}}+K_0^{(3),v}\notag\\
&+(-9.47414+0.574190n_l)\frac{\log(1-z)}{\sqrt{1-z}}\notag\\
&+(-17.5557+2.37068n_l-0.0690848n_l^2)\log(1-z)\notag\\
&+(1.31710+0.0312341n_l+0.00194703n_l^2)\log(1-z)^2\notag\\
&+(-0.630208+0.0763889n_l-0.00231481n_l^2)\log(1-z)^3\notag\\
&+{\cal O}(\sqrt{1-z})\displaybreak[0]\,,\notag\\
\Pi^{(3),p}(z)=&\frac{2.63641}{1-z}+\frac{-24.9710+0.678207n_l}{\sqrt{1-z}}+K_0^{(3),p}\notag\\
&+(-9.47414+0.574190n_l)\frac{\log(1-z)}{\sqrt{1-z}}\notag\\
&+(-10.9576+2.56218n_l-0.0690848n_l^2)\log(1-z)\notag\\
&+(3.23760+0.00345635n_l+0.00194703n_l^2)\log(1-z)^2\notag\\
&+(-0.630208+0.0763889n_l-0.00231481n_l^2)\log(1-z)^3\notag\\
&+{\cal O}(\sqrt{1-z})\displaybreak[0]\,,\notag\\
\Pi^{(3),a}(z)=&-0.731082\log(1-z)+K_0^{(3),a}+{\cal O}(\sqrt{1-z})\,,\notag\\
\Pi^{(3),s}(z)=&-1.09662\log(1-z)+K_0^{(3),s}+{\cal O}(\sqrt{1-z})\,,
\end{align}
where $n_l$ is the number of light quarks.
The corresponding analytic expressions and a brief outline of the
derivation of the expansions can be found in Appendix \ref{sec:thr_exp}.

The
construction of $\Pi^{(3),\delta}_{log}(z)$ is based on the expansions of
$\Pi^{(1),v}$ and $G(z)$ in the threshold region:
\begin{align}
  \label{eq:16}
  G(z) &= \frac{\pi}{2}\frac{1}{\sqrt{1-z}}+{\cal O}((1-z)^0)\,,\\
\Pi^{(1),v}(z)=&-\frac{3}{16}\log\left(1-z\right)+\text{const} +{\cal O}(\sqrt{1-z}) \,.
\end{align}
Obviously, we can obtain logarithms from $\Pi^{(1),v}(z)$ and powers of
$\frac{1}{\sqrt{1-z}}$ from $G(z)$.
Therefore, as a first attempt, we choose $\Pi^{(3),\delta}_{log}(z)$ to be a linear
combination of the form
\begin{equation}
  \label{eq:17}
  \Pi^{(3),\delta}_{log}(z) = \sum_{i>0,j} k_{ij} \Pi^{(1),v}(z)^i  G(z)^j\,.
\end{equation}
This choice of $\Pi^{(3),\delta}_{log}(z)$ will be modified later on
when we consider the high energy region.
The bounds of summation are chosen according 
to the powers of $\log(1-z)$ and  $\frac{1}{\sqrt{1-z}}$ that appear in
the threshold expansions of the polarisation functions. 
The condition that this ansatz reproduces the known logarithmic
contributions leads to a linear system of equations for the coefficients
$k_{ij}$, which determines them uniquely.

After taking care of the threshold logarithms, we treat those appearing
in the high energy expansions of the correlators\footnote{
In the formulas below we also give some not yet published
results. Namely, the first two non-logarithmic terms in the $1/z$  expansion of the 
scalar, pseudo-scalar and axial-vector correlators  come from Ref. \cite{Baikov:prel},
while the logarithmic contribution of order $1/z^2$ to the scalar and pseudo-scalar
correlators come from Ref. \cite{Chetyrkin:prel}.}: 
\begin{align}
  \label{eq:18}
\Pi^{(3),v}(z)=&-10.0036+1.37572n_l-0.0328147n_l^2\notag\\
&+(-0.357488+0.102421n_l-0.00218365n_l^2)\log(-4z)\notag\\
&+(0.193107-0.0254675n_l+0.000486744n_l^2)\log(-4z)^2\notag\\
&+(-0.0563482+0.00727073n_l-0.000234540n_l^2)\log(-4z)^3\notag\\
&+\big[-7.11044+1.01908n_l-0.0310950n_l^2\notag\\
&+(-5.88388+0.753052n_l-0.0146587n_l^2)\log(-4z)\notag\\
&+(2.82917-0.251016n_l+0.00457353n_l^2)\log(-4z)^2\notag\\
&+(-0.416015+0.0344773n_l-0.000703619n_l^2)\log(-4z)^3\big]\frac{1}{z}\notag\\
 &+\big[D_2^{(3),v}+(-7.85787+0.987298n_l-0.0260187n_l^2)\log(-4z)\notag\\
&+(0.215865+0.0367569n_l-0.000883940n_l^2)\log(-4z)^2\notag\\
&+(0.525948-0.0435071n_l+0.000674302n_l^2)\log(-4z)^3\notag\\
&+(-0.0955383+0.00589281n_l-0.0000879524n_l^2)\log(-4z)^4\big]\frac{1}{z^2}\notag\\
&+{\cal O}\left(\frac{1}{z^3}\right)\,,\notag\\
\Pi^{(3),p}(z)=&-25.1130+3.48518n_l-0.102852n_l^2\notag\\
&+(2.20686-0.230808n_l+0.0142957n_l^2)\log(-4z)\notag\\
&+(6.14249-0.637024n_l+0.0110499n_l^2)\log(-4z)^2\notag\\
&+(-1.56708+0.135388n_l-0.00257994n_l^2)\log(-4z)^3\notag\\
&+(0.104004-0.00861934n_l+0.000175905n_l^2)\log(-4z)^4\notag\\
&+\big[-1.35821+0.177211n_l-0.00711947n_l^2\notag\\
&+(-17.9226+2.10947n_l-0.0515675n_l^2)\log(-4z)\notag\\
&+(3.44902-0.144872n_l+0.00118877n_l^2)\log(-4z)^2\notag\\
&+(0.559407-0.0548237n_l+0.000820889n_l^2)\log(-4z)^3\notag\\
&+(-0.191077+0.0117856n_l-0.000175905n_l^2)\log(-4z)^4\big]\frac{1}{z}\notag\displaybreak[0]\\
&+\big[D_2^{(3),p}+(-7.80715+0.685314n_l-0.0148423n_l^2)\log(-4z)\notag\\
&+(-1.79679+0.226979n_l-0.00452222n_l^2)\log(-4z)^2\notag\\
&+(1.61678-0.107265n_l+0.00148053n_l^2)\log(-4z)^3\notag\\
&+(-0.184997+0.00989465n_l-0.000131929n_l^2)\log(-4z)^4\big]\frac{1}{z^2}\notag\\
&+{\cal O}\left(\frac{1}{z^3}\right)\,,\notag\\
\Pi^{(3),a}(z)=&-9.05417+1.17501n_l-0.0275071n_l^2\notag\\
&+(-0.357488+0.102421n_l-0.00218365n_l^2)\log(-4z)\notag\\
&+(0.193107-0.0254675n_l+0.000486744n_l^2)\log(-4z)^2\notag\\
&+(-0.0563482+0.00727073n_l-0.000234540n_l^2)\log(-4z)^3\notag\\
&+\big[15.8531-2.36372n_l+0.0700597n_l^2\notag\\
&+(-5.65395+0.850654n_l-0.0286846n_l^2)\log(-4z)\notag\\
&+(-3.72933+0.420485n_l-0.00717996n_l^2)\log(-4z)^2\notag\\
&+(1.15106-0.100911n_l+0.00187632n_l^2)\log(-4z)^3\notag\\
&+(-0.104004+0.00861934n_l-0.000175905n_l^2)\log(-4z)^4\big]\frac{1}{z}\notag\\
&+\big[D_2^{(3),a}+(6.89148-0.858786n_l+0.0213308n_l^2)\log(-4z)\notag\\
&+(-0.537351-0.00471280n_l+0.000272690n_l^2)\log(-4z)^2\notag\\
&+(-0.444548+0.0443867n_l-0.000850207n_l^2)\log(-4z)^3\notag\\
&+(0.0955383-0.00589281n_l+0.0000879524n_l^2)\log(-4z)^4\big]\frac{1}{z^2}\notag\\
&+{\cal O}\left(\frac{1}{z^3}\right)\,,\notag\\
\Pi^{(3),s}(z)=&-32.1410+4.42783n_l-0.125454n_l^2\notag\\
&+(2.20686-0.230808n_l+0.0142957n_l^2)\log(-4z)\notag\\
&+(6.14249-0.637024n_l+0.0110499n_l^2)\log(-4z)^2\notag\\
&+(-1.56708+0.135388n_l-0.00257994n_l^2)\log(-4z)^3\notag\\
&+(0.104004-0.00861934n_l+0.000175905n_l^2)\log(-4z)^4\notag\\
&+\big[40.4451-5.58423n_l+0.173022n_l^2\notag\\
&+(-27.2831+3.37357n_l-0.0984771n_l^2)\log(-4z)\notag\\
&+(-8.17872+0.943070n_l-0.0168387n_l^2)\log(-4z)^2\notag\\
&+(4.73545-0.353041n_l+0.00527714n_l^2)\log(-4z)^3\notag\\
&+(-0.573230+0.0353569n_l-0.000527714n_l^2)\log(-4z)^4\big]\frac{1}{z}\notag\\
&+\big[D_2^{(3),s}+(16.9813-1.84595n_l+0.0461425n_l^2)\log(-4z)\notag\\
&+(-5.04041+0.162549n_l+0.00179429n_l^2)\log(-4z)^2\notag\\
&+(-1.58168+0.153983n_l-0.00233074n_l^2)\log(-4z)^3\notag\\
&+(0.554991-0.0296839n_l+0.000395786n_l^2)\log(-4z)^4\big]\frac{1}{z^2}\notag\\
&+{\cal O}\left(\frac{1}{z^3}\right)\,.
\end{align}
The high energy logarithms can be generated through
$G(z)$:
\begin{equation}
  \label{eq:19}
  G(z) = \frac{-\log(-4z)}{2z}+{\cal O}\left(\frac{1}{z^2}\right)\,.
\end{equation}
However, 
the behaviour in the threshold region must not be spoiled by additional poles induced by the threshold
expansion of $G(z)$ (see Eq. \eqref{eq:16}). The
 simplest way to avoid this would be to multiply $G(z)$ by a factor
$\sqrt{1-z}$, which can however introduce non-integer powers of $z$ in
the high energy expansion. Following these
considerations we extend the ansatz \eqref{eq:17} to
\begin{equation}
  \label{eq:20}
  \Pi^{(3),\delta}_{log}(z) = \sum_{i>0,j} k_{ij} \Pi^{(1),v}(z)^i
  G(z)^j+\sum_{m,n} d_{mn} \big(z\,G(z)\big)^m \left(1-\frac{1}{z}\right)^{\left \lceil
      \frac{m}{2}\right \rceil} \frac{1}{z^n}\,,
\end{equation}
where $ \lceil\dots\rceil$ means rounding up to the next integer number.
The coefficients $d_{mn}$ with $m>0$ are again fixed by demanding  that
the known logarithms of $\Pi^{(3),\delta}$ are reproduced correctly up
to the highest known order. This leads to singular terms in the low
energy expansion of $\Pi^{(3),\delta}_{log}(z)$. We adjust the
coefficients $d_{0n}$ in such a way that these singular terms are
cancelled. Furthermore we can retain the property
$\Pi^{(3),\delta}_{reg}(0)=0$, which also means
 $\Pi^{(3),\delta}_{log}(0)=0$ by Eq. \eqref{eq:10}, by choosing the
 remaining free coefficient $d_{00}$ accordingly.

The choice of $\Pi^{(3),\delta}_{log}(z)$ is of course not unique. A perfect
reconstruction of the polarisation function would clearly not depend on
the specific choice, so variations of $\Pi^{(3),\delta}_{log}(z)$ can be
used to estimate the quality of the approximation procedure. Following
Ref. \cite{Hoang:2008qy}, we introduce additional parameters $a_i$ and $b_i$ for this purpose.
For the vector and pseudo-scalar correlators we modify the ansatz
Eq. \eqref{eq:20} for $\Pi^{(3),\delta}_{log}(z)$ in the following way:
In the first sum, we multiply the summands with  $i=3,\,j=0$  and
$i=j=1$ (which roughly correspond to terms proportional to  $\log^3
(1-z)$ and $\log(1-z)/\sqrt{1-z}$) by factors  $a_1+1/z$ and $a_2+1/z$,
respectively. In the second sum all summands corresponding to the two highest powers
of logarithms are multiplied by factors
$1+1/(b_1z)$ for $m=3$ and $1+1/(b_2z)$ for $m=4$.

In the cases of the axial-vector and scalar correlator the threshold
behaviour is quite different and only one term proportional to $\log
(1-z)$ is known. Consequently the first sum in the ansatz 
Eq. \eqref{eq:20} for $\Pi^{(3),\delta}_{log}(z)$ shrinks to a single term. This term is multiplied by $a_1+1/z$. In order to
arrive at a total number of four parameters like in the vector and
pseudo-scalar case, we multiply the terms corresponding to the three
highest powers of logarithms in the second sum by factors
$1+1/(b_1z)$, $1+1/(b_2z)$ and $1+1/(b_3z)$.

Except for the conditions $a_i \neq -1$ and $b_i \neq 0$ the values of
the parameters are in principle arbitrary. We vary them independently
with
\begin{align}
  \label{eq:21}
  a_i \in & \{-1 \pm 1,\, -1\pm 4,\, -1 \pm 16,\, -1 \pm 64\}\,,\notag\\
  b_i \in & \{\pm1,\, \pm 4,\, \pm 16,\, \pm64\}\,.
\end{align}

\subsection{Pad\'e approximation}
\label{sec:pade2}

In the next step we determine the coefficients of the Pad\'e approximants
from the expansions in the low energy, threshold, and euclidean high
energy region. The low energy expansions are taken from \cite{Maier:2009}; in numerical
form they read
\begin{align}
  \label{eq:22}
  \Pi^{(3),v}(z)=&(10.6103-1.30278n_l+0.0282783n_l^2)z\notag\\
  & +(10.4187-1.12407n_l+0.0223706n_l^2)z^2\notag\\
  & +(10.2031-1.01971n_l+0.0194021n_l^2)z^3+ {\cal O}(z^4)\notag\,,\displaybreak[0]\\
  \Pi^{(3),p}(z)=&(0.812723-0.190853n_l+0.00721861n_l^2)z\notag\\
  & +(6.33595-0.693155n_l+0.0145600n_l^2)z^2\notag\\
  & +(8.36076-0.803494n_l+0.0154075n_l^2)z^3\notag\\
  & +(9.14377-0.818646n_l+0.0149416n_l^2)z^4+ {\cal O}(z^5)\notag\,,\displaybreak[0]\\
  \Pi^{(3),a}(z)=&(4.84212-0.610731n_l+0.0141353n_l^2)z\notag\\
  & +(2.93924-0.335580n_l+0.00716845n_l^2)z^2\notag\\
  & +(2.06278-0.222971n_l+0.00461424n_l^2)z^3+ {\cal
    O}(z^4)\notag\,,\displaybreak[0]\\
\Pi^{(3),s}(z)=&(0.123690-0.0679839n_l+0.00455586n_l^2)z\notag\\
 &+(1.78515-0.232769n_l+0.00574404n_l^2)z^2\notag\\
 &+(1.92442-0.215014n_l+0.00469613n_l^2)z^3+ {\cal O}(z^4)\,.
\end{align}
The expansions around threshold and for high energies are listed in
Eqs. \eqref{eq:15} and \eqref{eq:18}, respectively.

Following Ref. \cite{Hoang:2008qy}, we additionally require that terms proportional to
$z^{-\frac{3}{2}}$ and $z^{-\frac{5}{2}}$ are absent in the high energy
expansion. The resulting number of constraints from the different kinematic
regions is listed in Table \ref{tab:I}.

\begin{table}
  \begin{center}
    \begin{tabular}{|c|c|c|c|c|}
      \hline
      current      & low energy & threshold & high energy & total\\\hline
      vector       & 3          & 2         & 2+2         &9     \\
      axial-vector & 3          & 0         & 2+2         &7     \\
      scalar       & 3          & 0         & 2+2         &7     \\
      pseudo-scalar & 4          & 2         & 2+2         &10    \\\hline
    \end{tabular}
  \end{center}

  \caption{Number of constraints from the various kinematical regions
    for the different current correlators. In the high energy region
    constraints from both known expansion coefficients and absence of
    terms with half-integer powers of $1/z$ are listed}

  \label{tab:I}
\end{table}

\subsection{Discussion of approximants}
\label{sec:disc-appr}

It turns out that some of the approximants have poles inside the unit
circle which translate to unphysical poles inside the complex $z$-plane for the
reconstructed polarisation functions. For this reason we immediately discard
approximants which have poles for $|\omega|<1$.

There is an additional subtlety for poles which do not lie inside, but
close to the unit circle. As pointed out in Ref. \cite{Hoang:2008qy}, these poles can have a
huge numerical effect on the behaviour of the approximant on the perimeter of the
circle. More specifically, they can lead to unphysical peaks in the
cross sections derived from the imaginary parts of the polarisation
functions above threshold. 

To avoid such resonances we discard approximants 
that have pronounced maxima on the perimeter of the circle, i.e.
\begin{equation}
  \label{eq:24}
  \max \bigg| p_{n,m}(\omega)\big|_{|\omega|=1}\bigg|>\rho\,,
\end{equation}
where the value of
$\rho$ is chosen heuristically using the following two criteria:
first, the real and imaginary parts of the polarisation functions should
show no significant additional peaks; second, an adequate number of
${\cal O}(1000)$ to ${\cal O}(10000)$ ``good'' approximants should remain.
In practice, we choose $\rho=3$ for the vector, scalar and pseudo-scalar
correlators and $\rho=1.2$ for the axial-vector correlator. 

\section{Results}
\label{sec:res}

\begin{figure}
  \centering
  \begin{tabular}{cc}
    \includegraphics[width=.47\linewidth]{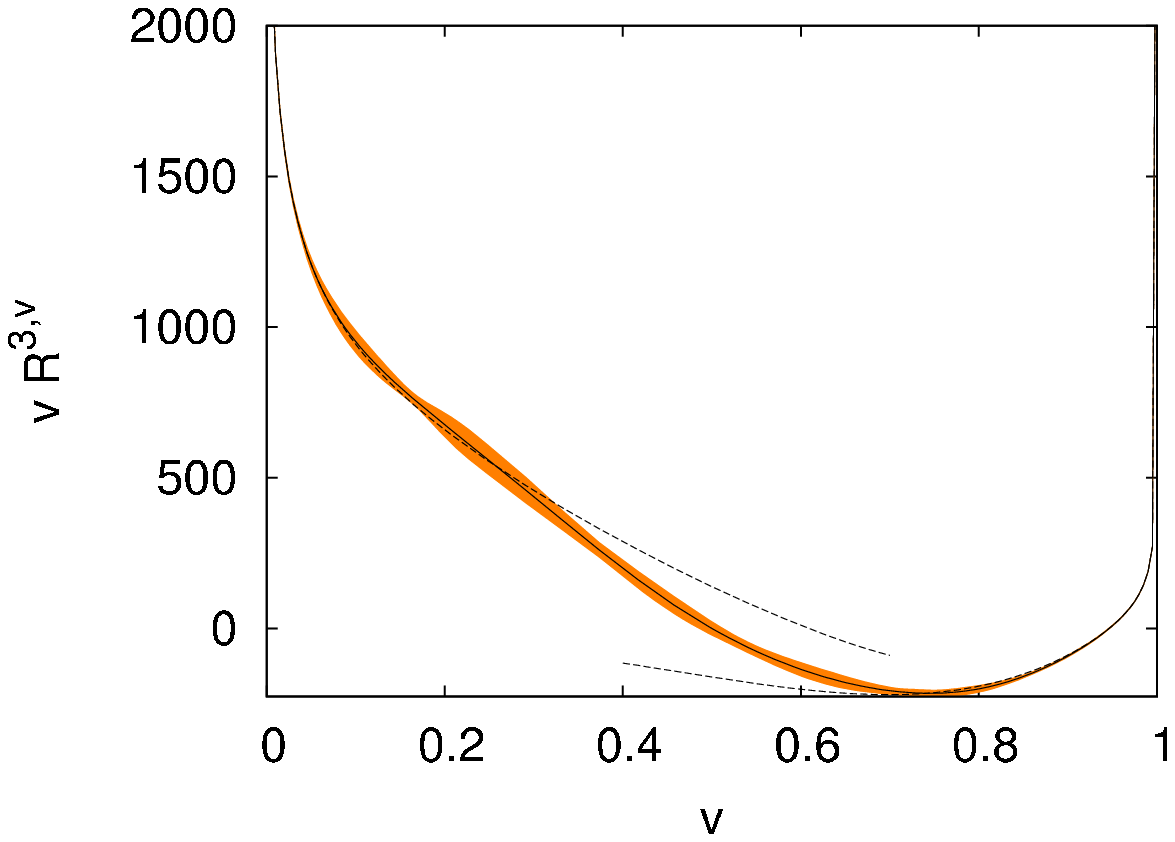}
&\includegraphics[width=.47\linewidth]{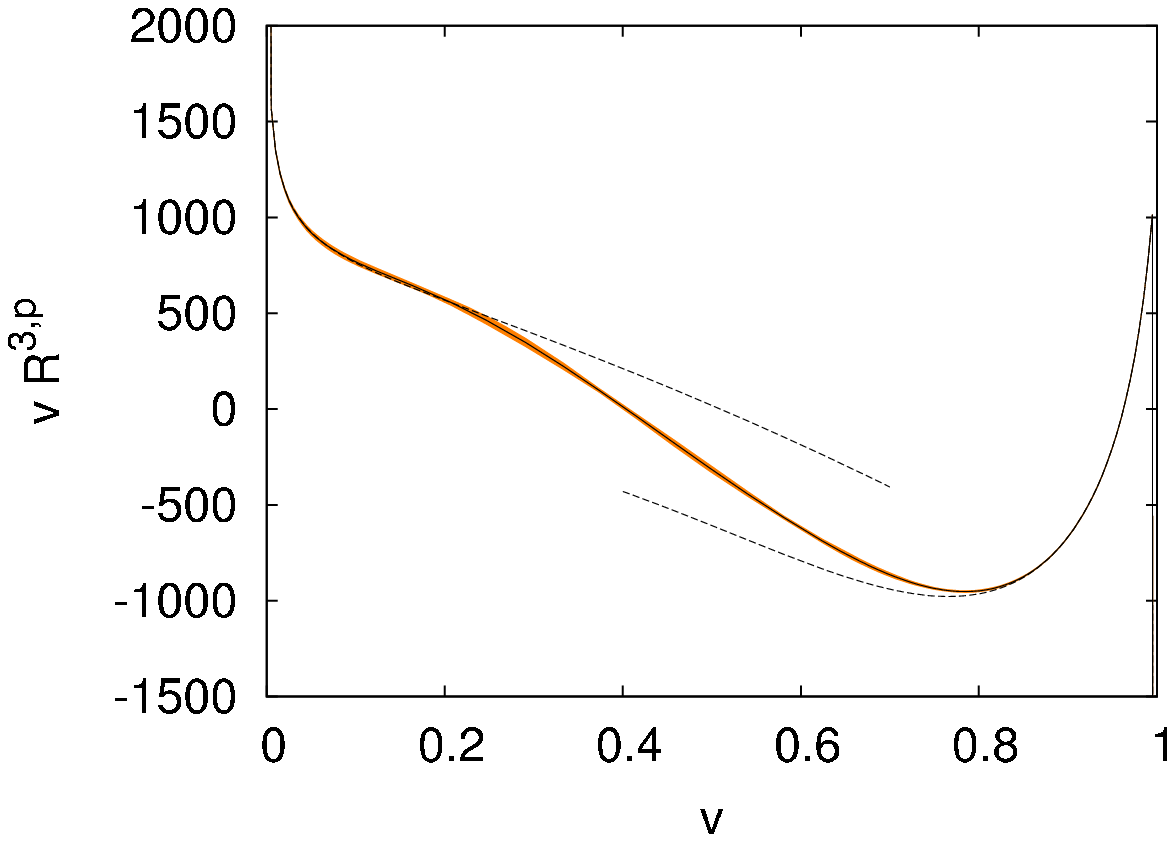}\\
\includegraphics[width=.47\linewidth]{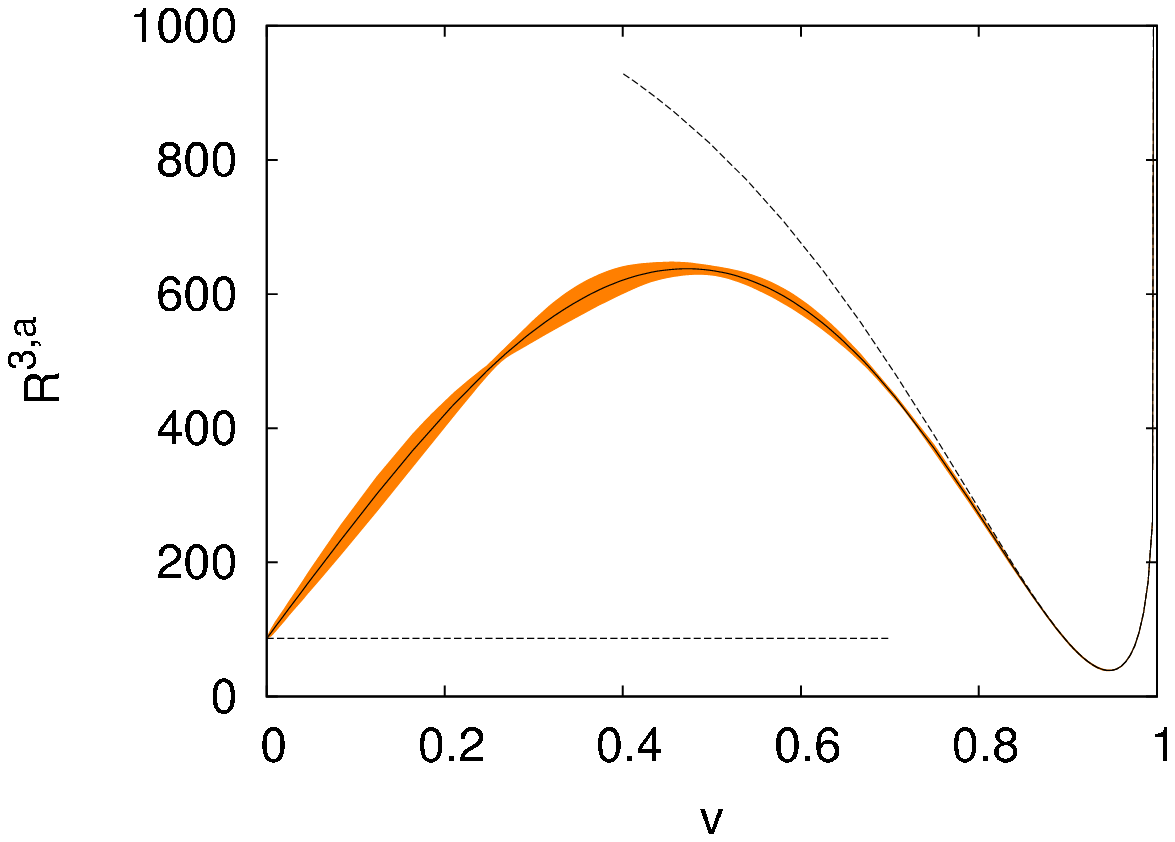}
&\includegraphics[width=.47\linewidth]{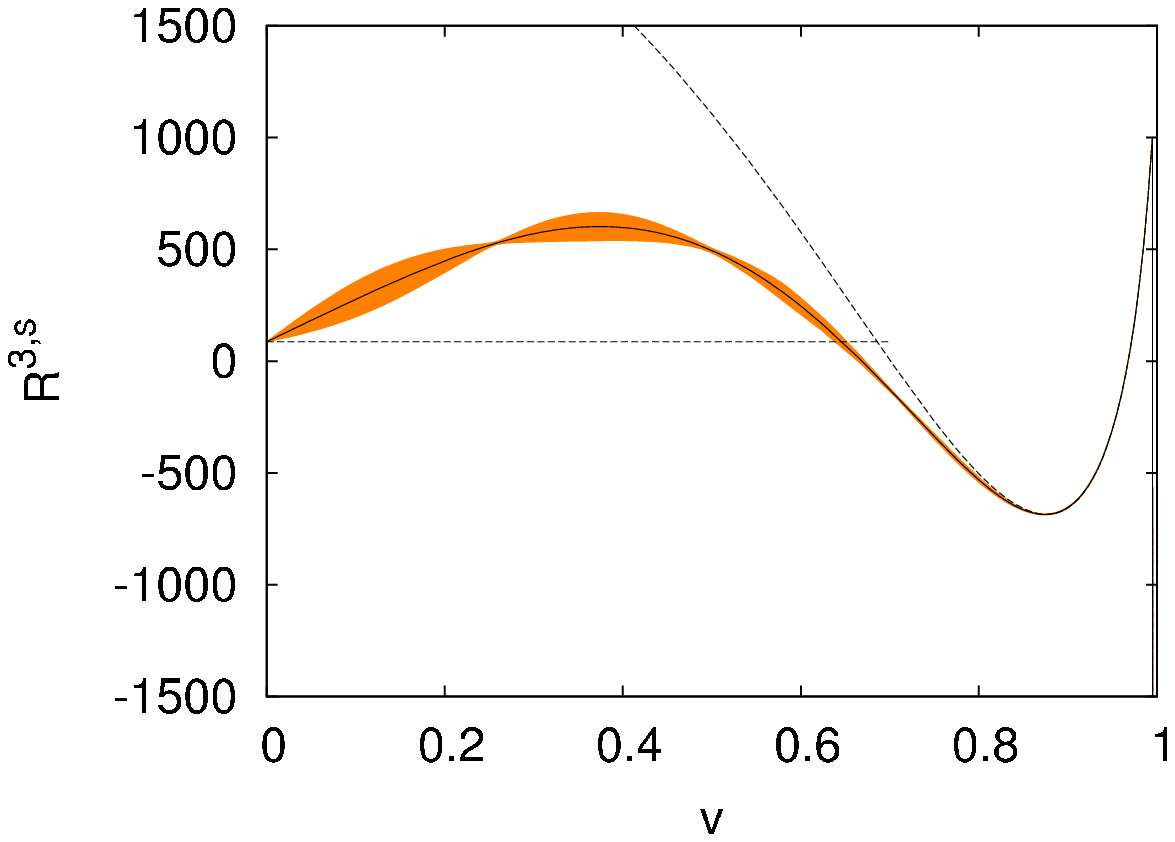}
\\
  \end{tabular}
  \caption{Imaginary part of the four loop contributions to the vector, pseudo\--scalar,
    axial-vector  and  scalar polarisation
    functions above the charm threshold. The plots show $v
    R^{(3),v}=v 12\pi \text{Im}(\Pi^{(3),v})$,
$v R^{(3),p}=v 8\pi \text{Im}(\Pi^{(3),p})$, $R^{(3),a}=12\pi
    \text{Im}(\Pi^{(3),a})$  and $R^{(3),s}=8\pi \text{Im}(\Pi^{(3),s})$
    as functions of $v=\sqrt{1-1/z}$. The solid black line is the
    mean from all approximants, the area covered by three standard
    deviations is shown by bands. The dashed lines show the
    expansions in the threshold and high energy regions (see
    Eqs. \eqref{eq:15} and \eqref{eq:18}).}
  \label{fig:2}
\end{figure}

\begin{figure}
  \centering
   \begin{tabular}{cc}
    \includegraphics[width=.47\linewidth]{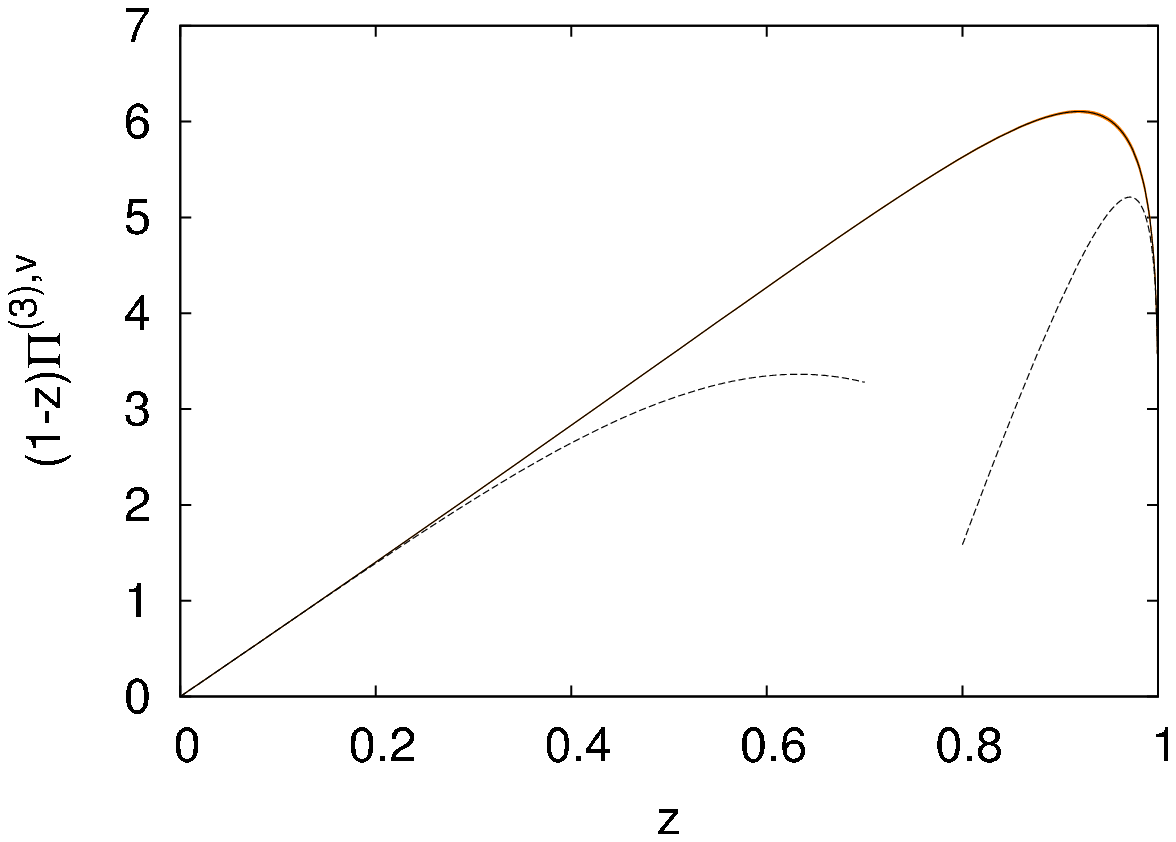}&\includegraphics[width=.47\linewidth]{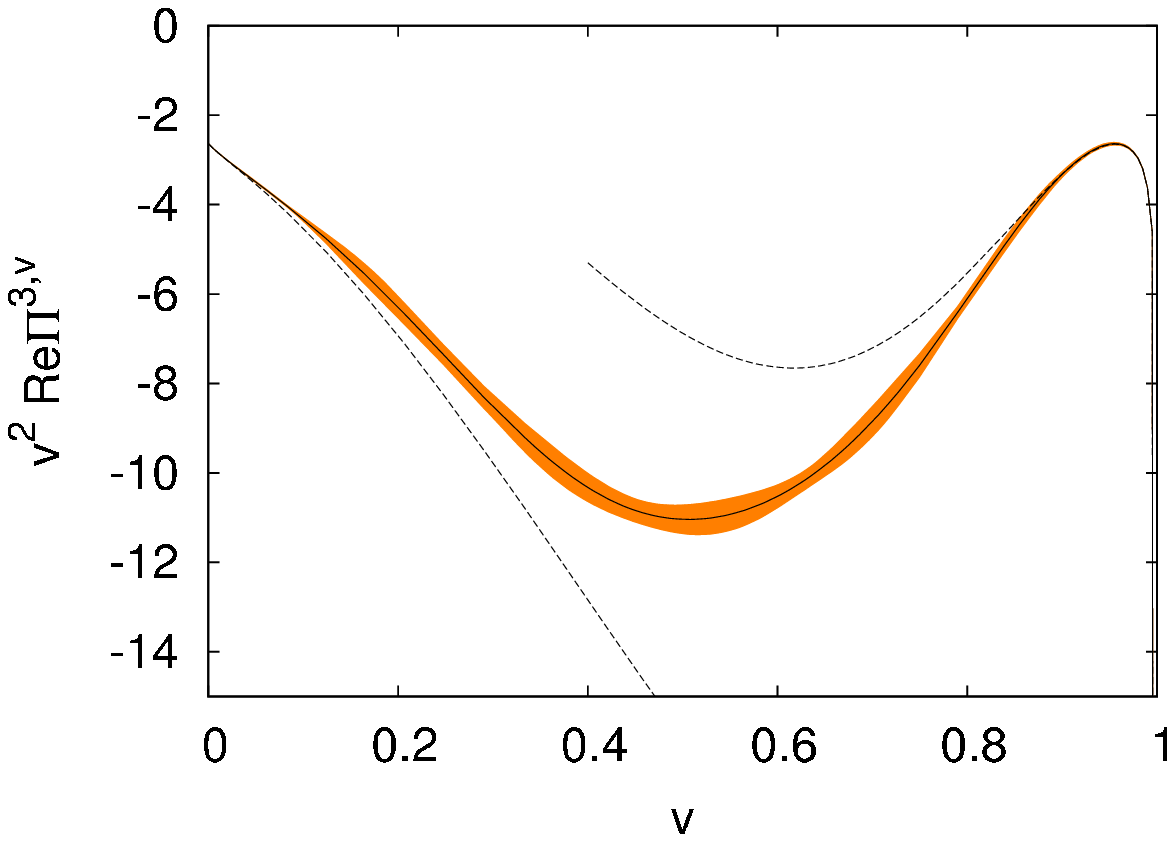}
  \end{tabular}
  \caption{Real part of the four loop contribution to the vector
    polarisation function for the case of charm quarks. On the left hand side the
  region below threshold is shown, on the right hand side the behaviour
  above threshold is plotted as a function of $v=\sqrt{1-1/z}$. The dashed lines show
  the known expansions in the respective regions (see Eqs. \eqref{eq:15} and
  \eqref{eq:22}), the solid lines are
  the mean values of all approximants. The shaded areas show the
  variation given by three standard deviations. In order to obtain finite
  values at threshold, $\text{Re}\,\Pi^{(3),v}$ is plotted with an extra factor
$1-z$ below and a factor $v^2$ above threshold.}
  \label{fig:3}
\end{figure}

From the Pad\'e approximants we reconstruct the polarisation
functions using the corresponding equations \eqref{eq:10}, \eqref{eq:11}
and \eqref{eq:12}. Their imaginary parts corresponding to hadron
production cross sections are plotted above the charm threshold
(i.e. with $n_l=3$) in
Fig. \ref{fig:2} for all four correlators. The real parts are less
important from a phenomenological point of view; as an example the vector polarisation function below and
above the charm threshold is shown in Fig. \ref{fig:3}. The error in the
low energy region turns out to be remarkably small.
Admittedly, this is  not unexpected considering the fact that the polarisation
functions are analytical in this region and that there is plenty of
information from low energy coefficients.

A selection of ``typical'' Pad\'e approximants for all four correlators
with $n_l=3,\ 4$ and  $5$ massless quarks can be downloaded from \\{\tt http://www-ttp.particle.uni-karlsruhe.de/Progdata/ttp09/ttp09-17/}

\begin{figure}
  \centering
  \includegraphics{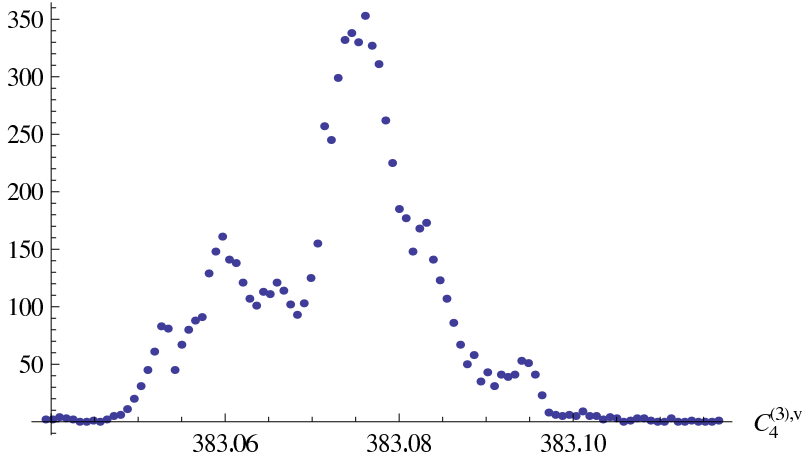}
  \caption{Distribution of the values of $C_4^{(3),v}$ in on-shell
    scheme from different Pad\'e approximants to
    the charm vector correlator.}
  \label{fig:4}
\end{figure}

The reconstructed functions can be expanded again to obtain additional
low energy, threshold and high energy coefficients. We find that the
values of the coefficients are strongly peaked around the mean
value (for an example, see Fig. \ref{fig:4}). As a consequence we give our errors in terms of standard
deviations. 
As expected 
the error is a lot
smaller for the low energy coefficients in comparison to the coefficients
in the threshold and high energy regions. 

There are very few coefficients in the threshold and high energy regions
which differ from the mean value by more than 50 standard deviations. We
discard these coefficients. The resulting
estimates for the expansion coefficients are shown in Tables
\ref{tab:2} to \ref{tab:5}. 

Comparing our  predictions for the vector correlator with the previous results \cite{Hoang:2008qy} (see Table \ref{tab:6}), we find indications for a
different sign of the threshold constant $K^{(3),v}_0$, but good agreement for all low
energy moments. Furthermore, the errors
are decreased notably, depending on the coefficient by about one order
of magnitude. This is mainly due to the additional information from
 $C^{(3),v}_3$, whereas $D^{(3),v}_0$ and $D^{(3),v}_1$ seem to play
 only a minor role. 
The impact of additional information on
the quality of the predictions from the Pad\'e approximants can also be
seen in Table \ref{tab:7}, where we compare results for the pseudo-scalar
correlator with three and four low energy coefficients as input.

\begin{table}
  \centering
\fontsize{9}{12}\selectfont
  \begin{tabular}{cc}
$\begin{array}{|c|c|c|c|}
\hline
                    &   n_l=3     & n_l=4       & n_l=5       \\\hline
C^{(3),v}_1          & 366.1748    & 308.0188    & 252.8399       \\\hline
C^{(3),v}_2          & 381.5091    & 330.5835    & 282.0129       \\\hline
C^{(3),v}_3          & 385.2331    & 338.7065    & 294.2224      \\\hline
C^{(3),v}_4          & 383.073(11) & 339.913(10) & 298.576(9)    \\\hline
C^{(3),v}_5          & 378.688(32) & 338.233(32) & 299.433(27)   \\\hline
C^{(3),v}_6          & 373.536(61) & 335.320(63) & 298.622(54)  \\\hline
C^{(3),v}_7          & 368.23(9)   & 331.90(10)  & 296.99(9)    \\\hline
C^{(3),v}_8          & 363.03(13)  & 328.33(14)  & 294.94(12)   \\\hline
C^{(3),v}_9          & 358.06(17)  & 324.78(18)  & 292.72(16)   \\\hline
C^{(3),v}_{10}       & 353.35(20)  & 321.31(22)  & 290.44(19)   \\\hline
K^{(3),v}_0          &   17(11)    &  17(29)     &  16(10)     \\\hline
D^{(3),v}_2          & 2.0(42)     & 1.2(83)     & 1.4(21)    \\\hline
  \end{array}$
&
$\begin{array}{|c|c|c|c|}
\hline
              & n_l=3        & n_l=4        & n_l=5        \\\hline
C^{(3),p}_1    & 16.0615      & 8.6753       & 2.0489       \\\hline
C^{(3),p}_2    & 230.9502     & 199.8289     & 170.2403     \\\hline
C^{(3),p}_3    & 320.5093     & 283.8922     & 248.8971     \\\hline
C^{(3),p}_4    & 359.1116     & 321.5253     & 285.5120     \\\hline
C^{(3),p}_5    & 376.3673(23) & 339.2386(21) & 303.6025(20) \\\hline
C^{(3),p}_6    & 383.6206(84) & 347.4338(75) & 312.6556(70) \\\hline
C^{(3),p}_7    & 385.794(18)  & 350.695(17)  & 316.925(16)  \\\hline
C^{(3),p}_8    & 385.250(32)  & 351.252(29)  & 318.511(28)  \\\hline
C^{(3),p}_9    & 383.215(48)  & 350.278(44)  & 318.533(42)  \\\hline
C^{(3),p}_{10} & 380.360(66)  & 348.424(61)  & 317.620(58)  \\\hline
K^{(3),p}_0    &  2(76)       &  8(27)       &  11(42)      \\\hline
D^{(3),p}_2    & 4.98(57)     &  4.11(48)    &  3.46(45)   \\\hline
  \end{array}$
\end{tabular}
\caption{Expansion coefficients from the reconstructed vector and
    pseudo-scalar polarisation functions for different numbers of light
    quarks in the on shell scheme. $C^{(3),v}_{1-3}$ and $C^{(3),p}_{1-4} $ are known exactly. The errors always apply to the last digits, i.e. $2.0(42)$
    means an error of $4.2$.}
  \label{tab:2}
\end{table}
\begin{table}
  \centering

\fontsize{9}{12}\selectfont
  \begin{tabular}{cc}
$\begin{array}{|c|c|c|c|}
\hline
              & n_l=3   & n_l=4       & n_l=5       \\\hline
C^{(3),a}_1    &165.1328      &138.1938     & 112.7427     \\\hline
C^{(3),a}_2    &105.1185      & 90.0956     & 75.8274     \\\hline
C^{(3),a}_3    & 75.5564      & 65.5198     & 55.9690     \\\hline
C^{(3),a}_4    & 57.7298(29)  & 50.4287(42) & 43.4720(39) \\\hline
C^{(3),a}_5    & 46.005(9)    & 40.397(13)  & 35.048(12)  \\\hline
C^{(3),a}_6    & 37.813(17)   & 33.338(24)  & 29.065(22)  \\\hline
C^{(3),a}_7    & 31.825(25)   & 28.151(36)  & 24.639(33)  \\\hline
C^{(3),a}_8    & 27.291(34)   & 24.206(48)  & 21.255(44)  \\\hline
C^{(3),a}_9    & 23.759(42)   & 21.123(59)  & 18.599(54)  \\\hline
C^{(3),a}_{10} & 20.943(49)   & 18.658(69)  & 16.468(63)  \\\hline
K^{(3),a}_0    & 16.68(25)    & 14.28(48)   &  11.91(37)  \\\hline
D^{(3),a}_2    & 1.69(27)     &  1.26(38)   &  0.83(34)   \\\hline
  \end{array}$
&

  $\begin{array}{|c|c|c|c|}
\hline
              & n_l=3       & n_l=4       & n_l=5       \\\hline
C^{(3),s}_1    & -2.0665     & -3.9663     & -5.3866      \\\hline
C^{(3),s}_2    & 59.9301     & 49.7941     & 40.2628      \\\hline
C^{(3),s}_3    & 69.5687     & 59.9811     & 50.8880      \\\hline
C^{(3),s}_4    & 64.641(14)  & 56.534(14)  & 48.819(13)  \\\hline
C^{(3),s}_5    & 57.168(43)  & 50.399(41)  & 43.946(39)  \\\hline
C^{(3),s}_6    & 50.069(81)  & 44.374(76)  & 38.941(73)  \\\hline
C^{(3),s}_7    & 43.95(12)   & 39.10(12)   & 34.47(111)  \\\hline
C^{(3),s}_8    & 38.81(16)   & 34.64(15)   & 30.65(149)  \\\hline
C^{(3),s}_9    & 34.52(20)   & 30.89(19)   & 27.41(184)  \\\hline
C^{(3),s}_{10} & 30.93(24)   & 27.73(22)   & 24.67(216)  \\\hline
K^{(3),s}_0    &  17.4(11)   &  14.9(11)   & 12.6(12)    \\\hline
D^{(3),s}_2    &  7.7(10)    &  6.3(9)     & 5.1(8)      \\\hline
  \end{array}$
\end{tabular}
  \caption{Expansion coefficients from the reconstructed axial-vector and
    scalar polarisation functions for different numbers of light
    quarks in the on shell scheme. The coefficients $C^{(3),a}_{1-3}$ and
    $C^{(3),s}_{1-3}$ are known exactly.}
  \label{tab:3}
\end{table}

\begin{table}
  \centering
\fontsize{9}{12}\selectfont
  \begin{tabular}{cc}
  $\begin{array}{|c|c|c|c|}
\hline
                    &   n_l=3     & n_l=4       & n_l=5       \\\hline
\bar{C}^{(3),v}_1    & -5.6404     & -7.7624     & -9.6923       \\\hline
\bar{C}^{(3),v}_2    & -3.4937     & -2.6438     & -1.8258       \\\hline
\bar{C}^{(3),v}_3    & -2.8395     & -1.1745     & 0.4113       \\\hline
\bar{C}^{(3),v}_4    & -3.349(11)  & -1.386(10)  & 0.471(9)    \\\hline
\bar{C}^{(3),v}_5    & -3.737(32)  & -1.754(32)  & 0.104(27)   \\\hline
\bar{C}^{(3),v}_6    & -3.735(61)  & -1.910(63)  & -0.228(54)  \\\hline
\bar{C}^{(3),v}_7    & -3.39(10)   & -1.85(10)   & -0.46(9)    \\\hline
\bar{C}^{(3),v}_8    & -2.85(13)   & -1.67(14)   & -0.66(12)   \\\hline
\bar{C}^{(3),v}_9    & -2.22(17)   & -1.47(18)   & -0.91(16)   \\\hline
\bar{C}^{(3),v}_{10} & -1.65(20)   & -1.37(22)   & -1.30(19)   \\\hline
  \end{array}$
&
$\begin{array}{|c|c|c|c|}
\hline
                    & n_l=3        & n_l=4        & n_l=5        \\\hline
\bar{C}^{(3),p}_1    & -1.2224      & -7.2260      &-12.4695      \\\hline
\bar{C}^{(3),p}_2    &  7.0659      &  6.0605      &  5.1954     \\\hline
\bar{C}^{(3),p}_3    & 14.5789      & 14.8438      & 15.1394      \\\hline
\bar{C}^{(3),p}_4    & 13.3278      & 14.3313      & 15.3164      \\\hline
\bar{C}^{(3),p}_5    &  9.9948(23)  & 11.4153(21)  & 12.7852(19)  \\\hline
\bar{C}^{(3),p}_6    &  6.8011(84)  &  8.3991(75)  &  9.9221(70)  \\\hline
\bar{C}^{(3),p}_7    &  4.311(18)   &  5.907(17)   &  7.408(16)   \\\hline
\bar{C}^{(3),p}_8    &  2.548(32)   &  4.008(29)   &  5.356(28)   \\\hline
\bar{C}^{(3),p}_9    &  1.373(48)   &  2.594(44)   &  3.690(42)   \\\hline
\bar{C}^{(3),p}_{10} &  0.612(66)   &  1.517(61)   &  2.285(58)   \\\hline
  \end{array}$
\end{tabular}
  \caption{ Coefficients from the low energy expansion of the reconstructed vector and
    pseudo-scalar polarisation functions for different numbers of light
    quarks. The  coefficients are given in the
    $\overline{\text{MS}}$ scheme with
    $\mu=m$. $\bar{C}^{(3),v}_{1-3}$ and $\bar{C}^{(3),p}_{1-4}$ are known
    exactly.}
  \label{tab:4}
\end{table}

\begin{table}
  \centering
\fontsize{9}{12}\selectfont
  \begin{tabular}{cc}
$\begin{array}{|c|c|c|c|}
\hline
                    & n_l=3      & n_l=4      & n_l=5        \\\hline
\bar{C}^{(3),a}_1    & -2.4297    & -2.6606    & -2.7958      \\\hline
\bar{C}^{(3),a}_2    & -3.8059    & -2.8384    & -1.9120      \\\hline
\bar{C}^{(3),a}_3    & -2.6066    & -1.7770    & -0.9920      \\\hline
\bar{C}^{(3),a}_4    & -1.7688(29)& -1.1387(42)& -0.5498(39)  \\\hline
\bar{C}^{(3),a}_5    & -1.144(9)  & -0.692(13) & -0.278(12) \\\hline
\bar{C}^{(3),a}_6    & -0.678(17) & -0.376(24) & -0.109(22) \\\hline
\bar{C}^{(3),a}_7    & -0.344(25) & -0.166(36) & -0.022(33) \\\hline
\bar{C}^{(3),a}_8    & -0.121(34) & -0.047(48) & -0.004(44) \\\hline
\bar{C}^{(3),a}_9    &  0.009(42) & -0.004(59) & -0.046(54) \\\hline
\bar{C}^{(3),a}_{10} &  0.061(49) & -0.024(69) & -0.138(63)  \\\hline
  \end{array}$
&
  $\begin{array}{|c|c|c|c|}
\hline
                    & n_l=3       & n_l=4       & n_l=5       \\\hline
\bar{C}^{(3),s}_1    & -5.4135     & -7.0456     & -8.1981      \\\hline
\bar{C}^{(3),s}_2    &-12.9598     &-11.6485     &-10.3292      \\\hline
\bar{C}^{(3),s}_3    & -6.6011     & -5.4063     & -4.2477      \\\hline
\bar{C}^{(3),s}_4    & -3.972(14)  & -3.002(14)  & -2.073(13)  \\\hline
\bar{C}^{(3),s}_5    & -2.665(43)  & -1.903(41)  & -1.181(39)  \\\hline
\bar{C}^{(3),s}_6    & -1.843(81)  & -1.265(76)  & -0.723(73)  \\\hline
\bar{C}^{(3),s}_7    & -1.26(12)   & -0.84(12)   & -0.45(11)   \\\hline
\bar{C}^{(3),s}_8    & -0.82(16)   & -0.54(15)   & -0.29(15)   \\\hline
\bar{C}^{(3),s}_9    & -0.51(20)   & -0.35(19)   & -0.22(18)   \\\hline
\bar{C}^{(3),s}_{10} & -0.30(24)   & -0.25(22)   & -0.22(22)   \\\hline
  \end{array}$
\end{tabular}
  \caption{Coefficients from the low energy expansion of the reconstructed axial-vector and
    scalar polarisation functions for different numbers of light
    quarks. The  coefficients  are given in the
    $\overline{\text{MS}}$ scheme with $\mu=m$. The coefficients $\bar{C}^{(3),a}_{1-3}$ and
    $\bar{C}^{(3),s}_{1-3}$ are known exactly.}
  \label{tab:5}
\end{table}

\begin{table}
  \centering
  $\begin{array}{|l|c|c|c|c|c|c|}
\hline
      & \bar{C}^{(3),v}_3& \bar{C}^{(3),v}_4 & \bar{C}^{(3),v}_5 & \bar{C}^{(3),v}_6 &
     \bar{C}^{(3),v}_7 & K^{(3),v}_0 \\\hline
\text{Ref. \cite{Hoang:2008qy}}     & -3.28\pm 0.57 & -4.2 \pm 1.2 & -5.0 \pm 1.7 & -5.3 \pm
2.0 & -5.2 \pm 2.3 & -10 \pm 11
     \\\hline
\text{this work}     &-2.840\text{ (exact)}&-3.349(11)&-3.737(32)  & -3.735(61) & -3.39(10) &17(11)\\\hline
   \end{array}$
  \caption{Comparison of the $\overline{\text{MS}}$ low energy
    coefficients of the charm vector
    correlator with the previous results from \protect{\cite{Hoang:2008qy}}. 
    In the second row, the maximum error is estimated, whereas
  the error in the third row is given in terms of standard deviations.}
  \label{tab:6}
\end{table}

\begin{table}
  \centering
  $\begin{array}{|c|c|c|c|c|c|c|c|}
\hline
\text{input} & \bar{C}^{(3),p}_4& \bar{C}^{(3),p}_5 & \bar{C}^{(3),p}_6 & \bar{C}^{(3),p}_7 &
    \bar{C}^{(3),p}_8 & 
     K^{(3),p}_0 &  D^{(3),p}_2\\\hline
\bar{C}^{(3),p}_{1\text{-}3} &13.3310(91)&10.0053(286)&6.8221(567)   & 4.345(90)
& 2.596(128) 
& 4(169)& 4.98(63)
     \\\hline
\bar{C}^{(3),p}_{1\text{-}4} &13.3278  &9.9948(23)& 6.8011(84) &
4.311(18) & 2.548(32) 
& 2(76)& 4.98(57)\\\hline
   \end{array}$
  \caption{Comparison of results from different numbers of low energy
    coefficients. Shown are the predictions for the
    $\overline{\text{MS}}$ low energy coefficients of the $n_l=3$ pseudo-scalar
     polarisation function with three and four moments used as
    input.}
  \label{tab:7}
\end{table}

\section{Conclusion}
\label{sec:conc}
We have used the Pad\'e approximation method to reconstruct the full
energy dependence of heavy quark correlators for vector, axial-vector, 
scalar and pseudo-scalar currents at order $\alpha_s^3$. As input we have
used information from expansions in the low energy, threshold and high energy regions.
Expanding the reconstructed correlators, we have obtained predictions for additional
coefficients in these expansions. 
We find that these predictions are fairly accurate for low energy coefficients
but less precise for the coefficients in the threshold and high energy
expansions.

\section*{Acknowledgements}
\label{sec:ack}

We thank K.~G.~Chetyrkin, R.~V.~Harlander, J.~H.~K\"uhn,
C.~Rei\ss{}er and M. Steinhauser for helpful
discussions and carefully reading the manuscript. We are very grateful to P.~A.~Baikov, K.~G.~Chetyrkin,
R.~V.~Harlander and J.~H.~K\"uhn for sharing their results with us
prior to publication. We also thank A. Hoang and V. Mateu for useful
correspondence concerning their work \cite{Hoang:2008qy}.
This work was supported by the Deutsche Forschungsgemeinschaft through
the SFB/TR-9 ``Computational Particle Physics''. Ph.~M. was supported
by the Graduiertenkolleg ``Hochenergiephysik und Teilchenastrophysik''.
A.~M. thanks the Landesgraduiertenf\"orderung for support.

\appendix

\section{Threshold expansion}
\label{sec:thr_exp}

In this Appendix we summarise the threshold behaviour of 
all polarisation functions used in our paper. 
We derive the threshold behaviour of $\Pi^{(3),\delta}$ using 
Non-Relativistic QCD (NRQCD) \cite{Caswell:1985ui,Bodwin:1994jh}. For the 
present paper it is sufficient to obtain 
the results in an expansion in $(1-z)$ up to (and including) ${\cal O}((1-z)^0)$. 
To this end we need the matching coefficients
$c_\delta$ for the relation  $j^\delta=c_\delta\, j^\delta_{\rm NR}$ 
between QCD and NRQCD up to ${\cal O}(\alpha_s^2)$ for vector and
pseudo-scalar correlators, but only up to ${\cal O}(\alpha_s^0)$  for axial-vector 
and scalar correlators. The ${\cal O}(\alpha_s^2)$ matching coefficients
are derived in Ref. \cite{Czarnecki:1997vz,Beneke:1997jm} for the vector current (see also
Refs. \cite{Marquard:2006qi,Marquard:2009bj}), and in 
Ref. \cite{Kniehl:2006qw} for the axial-vector current.
The one-loop matching coefficients are known since long from standard QCD
computation (see e.g. \cite{Kniehl:2006qw} and references therein).
Heavy quark current correlators reduce to the correlators 
expressed in terms of NRQCD currents,
\begin{align}
i\int dx\, e^{iqx}\langle 0|\,{\rm T}j^{\delta}(x)j^{\delta}(0)|0\rangle
=
c_{\delta}^2\,
i\int dx\, e^{iqx}
\langle 0|\,{\rm T}j_{\rm NR}^{\delta}(x)j_{\rm NR}^{\delta}(0)|0\rangle
+ C_{\delta\delta}(q^2).
\label{eq:a1}
\end{align}
The second term on the right-hand side is an analytic function of
$q^2$, which corresponds to the hard heavy quark loop shrunk to a 
point from a diagrammatic point of view. 
This term does not contribute to $R(q^2)\propto{\rm Im}\Pi(q^2)$ due to its 
analyticity. For this reason it was  never calculated in NRQCD and the
constant $K_0^{(3),\delta}$ in the threshold expansion \eqref{eq:6}
remains unknown.

For  the vector case the calculation of the correlator in NRQCD was 
done analytically in Ref. \cite{Beneke:1999qg,Pineda:2006ri}. 
The expansion in $(1-z)$ can be easily obtained from it. 
For the other correlators we did not
find the explicit results in the literature, thus 
we performed the calculation for the present work and 
present the result in this Appendix. 

As an illustration we show several steps for the derivation
of the threshold expansion for the case of the vector correlator. Other
correlators can be obtained in a similar way.
The QCD vector current can be matched to the one in NRQCD by
\begin{align}
j^{v}_i(x)
=e^{2imx_0}\left(c_1+\frac{d_v}{6m}i\partial_0\right)
[\psi^\dag\sigma_i\chi](x),\qquad i=1,2,3\,,
\label{eq:a2}
\end{align}
where $c_1$ is the matching coefficient for the vector current whose
explicit form can be found in the references mentioned previously,
and $d_v=1$ at the order of interest. The time component of the 
vector current vanishes in the rest frame of the heavy quarks
with momentum $q=(q_0,\vec{0})$.
Substituting the QCD current by the NRQCD current the right-hand side
of Eq. \eqref{eq:a1} (neglecting $C_{\delta\delta}(q^2)$) is given by 
\begin{align}
\left(c_1-\frac{d_v}{6m}E\right)^2\,
i\,\int dx\, e^{iEx_0}
\langle 0| {\rm T}
[\chi^\dag\sigma_i\psi](x)\,
[\psi^\dag\sigma_j\chi](0)
|0 \rangle\,, 
\end{align}
where $E\equiv q_0-2m=\sqrt{q^2}-2m$ and integration by parts is used
to relate the derivative to $E$. It is well known that the correlators in NRQCD 
can be mapped onto Green's functions in quantum mechanics; thus one
obtains
\begin{align}
i\,\int dx\, e^{iEx_0}
\langle 0| {\rm T}
[\chi^\dag\sigma_i\psi](x)\,
[\psi^\dag\sigma_j\chi](0)
|0 \rangle 
=2\,N_c\, \delta_{ij} G(0,0;E),
\end{align}
where $N_c=3$ is the number of colours for QCD and $\delta_{ij}$ is
Kronecker's delta. The NNLO result for the Green's function
is presented in Eq. (A.29) of Ref. \cite{Pineda:2006ri}.
One should expand the (NNLL) Green's function presented 
in that reference and retain relevant terms of interest. Furthermore
UV divergences should be subtracted according to the $\overline{\rm MS}$
scheme conforming with the definition of $c_1$ in order to arrive at the
threshold expansion presented below.

Conventionally 
NRQCD computations are done using the effective coupling $\alpha_s^{(n_l)}(\mu)$ 
where the heavy quark is integrated out. 
We present our results using the pole mass $m$ and the QCD coupling constant
$\alpha_s^{(n_l+1)}(m)$ at the scale of the pole mass. 
Hence we re-express the effective coupling using the decoupling relation
\cite{Larin:1994va,Chetyrkin:1997un} 
\begin{eqnarray}
\alpha_s^{(n_l)}(m)=\alpha_s^{(n_l+1)}(m)
\left(1+d^{(2)} \bigg(\frac{\alpha_s^{(n_l+1)}(m)}{\pi}\bigg)^2 
+ {\cal O}(\alpha_s^3)\right),
\end{eqnarray}
with $d^{(2)}=-7/24$.
This induces a change in the vector and pseudo-scalar correlator  in
the coefficient of $\log(1-z)$ at order ${\cal O}((1-z)^0)$. 
The result for the vector and pseudo-scalar current correlators are given
by  
\begin{align}
\Pi^{(3),v}(z)=&\frac{2\pi^2\zeta_3}{9(1-z)}+\Big[-\frac{11\pi^3}{36}+\frac{\pi^3}{54}n_l\Big]\frac{\log(1-z)}{\sqrt{1-z}}\notag\\
&+\Big[-\frac{\pi^3}{108}-\frac{11\pi^3\log2}{18}-\frac{11\pi\zeta_3}{3}\notag\\
&\quad +n_l\Big(-\frac{5\pi^3}{162}+\frac{\pi^3\log2}{27}+\frac{2\pi\zeta_3}{9}\Big)\Big]\frac{1}{\sqrt{1-z}}\notag\displaybreak[0]\\
&+\Big[-\frac{121}{192}+\frac{11}{144}n_l-\frac{1}{432}n_l^2\Big]\log^3(1-z)\notag\displaybreak[0]\\
&+\Big[\frac{71}{96}+\frac{35\pi^2}{108}-\frac{121\log2}{32}+n_l\Big(-\frac{55}{192}+\frac{11\log2}{24}\Big)\notag\\
&\quad+n_l^2\Big(\frac{5}{432}-\frac{\log2}{72}\Big)\Big]\log^2(1-z)\notag\displaybreak[0]\\
&+\Big[-\frac{4169}{3456}-\frac{20741\pi^2}{5184}+\frac{9\pi^4}{256}+\frac{71\log2}{24}\notag\\
&\quad+\frac{56\pi^2\log2}{27}-\frac{121\log^22}{16}+\frac{1703\zeta_3}{288}\notag\\
&\quad+n_l\Big(\frac{79}{576}+\frac{11\pi^2}{48}-\frac{55\log2}{48}+\frac{11\log^22}{12}+\frac{13\zeta_3}{48}\Big)\notag\\
&\quad+n_l^2\Big(-\frac{25}{1296}-\frac{\pi^2}{144}+\frac{5\log2}{108}-\frac{\log^22}{36}\Big)\Big]\log(1-z)\notag\\
&+K^{(3),v}_0+{\cal O}(\sqrt{1-z})\,,\notag \displaybreak[0]
\\
\Pi^{(3),p}(z)=&\frac{2\pi^2\zeta_3}{9(1-z)}+\Big[-\frac{11\pi^3}{36}+\frac{\pi^3}{54}n_l\Big]\frac{\log(1-z)}{\sqrt{1-z}}\notag\\
&+\Big[\frac{7\pi^3}{108}-\frac{11\pi^3\log2}{18}-\frac{11\pi\zeta_3}{3}\notag\\
&\quad+n_l\Big(-\frac{5\pi^3}{162}+\frac{\pi^3\log2}{27}+\frac{2\pi\zeta_3}{9}\Big)\Big]\frac{1}{\sqrt{1-z}}\notag\displaybreak[0]\\
&+\Big[-\frac{121}{192}+\frac{11}{144}n_l-\frac{1}{432}n_l^2\Big]\log^3(1-z)\notag\displaybreak[0]\\
&+\Big[\frac{115}{96}+\frac{17\pi^2}{36}-\frac{121\log2}{32}+n_l\Big(-\frac{181}{576}+\frac{11\log2}{24}\Big)\notag\\
&\quad+n_l^2\Big(\frac{5}{432}-\frac{\log2}{72}\Big)\Big]\log^2(1-z)\notag\displaybreak[0]\\
&+\Big[-\frac{539}{128}-\frac{7207\pi^2}{1728}+\frac{9\pi^4}{256}+\frac{115\log2}{24}\notag\\
&\quad+3\pi^2\log2-\frac{121\log^22}{16}+\frac{287\zeta_3}{32}\notag\\
&\quad+n_l\Big(\frac{701}{1728}+\frac{11\pi^2}{48}-\frac{181\log2}{144}+\frac{11\log^22}{12}+\frac{13\zeta_3}{48}\Big)\notag\\
&\quad+n_l^2\Big(-\frac{25}{1296}-\frac{\pi^2}{144}+\frac{5\log2}{108}-\frac{\log^22}{36}\Big)\Big]\log(1-z)\notag\\
&+K^{(3),p}_0+{\cal O}(\sqrt{1-z})\notag\,.\\
\end{align}
Except for the term
proportional to
$\log(1-z)$ the result for the vector current agrees with Ref. \cite{Hoang:2008qy}\footnote{The overall numerical difference is very
  small. For $\alpha_s$ parametrised in terms of the ${\rm
    \overline{MS}}$ heavy quark mass,
we find full agreement with Ref.\cite{Hoang:2008qy}.}. The threshold behaviour for the
pseudo-scalar current is new.

The scalar and axial-vector correlators are suppressed by $(1-z)$
compared to those for vector and pseudo-scalar cases. However, their
leading term is divergent, and its $1/\epsilon$ pole generates a ${\rm log}(1-z)$  
in NRQCD. The result reads
\begin{align}
\label{s_a_thr}
\Pi^{(3),s}(z)=&-\frac{\pi^2}{9}\log(1-z)+K^{(3),s}_0+{\cal O}(\sqrt{1-z})\,,\notag\\
\Pi^{(3),a}(z)=&-\frac{2\pi^2}{27}\log(1-z)+K^{(3),a}_0+{\cal O}(\sqrt{1-z})\,.
\end{align}
The origin of the ${\rm log}(1-z)$ is a triple insertion of Coulomb
gluons which is of order ${\cal O}(\alpha_s^3)$ in NRQCD. The results in
Eq. \eqref{s_a_thr} are in
agreement with the one in Ref. \cite{Penin:1998ik}.

\bibliographystyle{hep}
\bibliography{hep}

\end{document}